\documentclass[aps,prd,twocolumn,superscriptaddress,showpacs,showkeys,nofootinbib]{revtex4}
\usepackage{graphicx}
\usepackage{euscript,amssymb}
\usepackage{amsfonts}
\usepackage{amsmath}
\usepackage{amssymb}

\begin{document}

\title{Worse than a big rip?}

\author{Mariam Bouhmadi-L\'{o}pez}
\email{mariam.bouhmadi@fisica.ist.utl.pt}
\affiliation{Centro Multidisciplinar de Astrof\'{\i}sica - CENTRA, Departamento de F\'{\i}sica, Instituto Superior T\'ecnico, Av. Rovisco Pais 1,
1049-001 Lisboa, Portugal}
\affiliation{Departamento de F\'{\i}sica,
Universidade da Beira Interior, R. Marqu\^{e}s d'\'Avila e Bolama,
6201-001 Covilh\~{a}, Portugal}
\affiliation{Institute of
Cosmology and Gravitation, University of Portsmouth,  Mercantile
House, Hampshire Terrace,  Portsmouth  PO1 2EG, UK}

\author{Pedro F. Gonz\'{a}lez-D\'{\i}az}
\email{p.gonzalezdiaz@imaff.cfmac.csic.es}
\author{Prado Mart\'{\i}n-Moruno}
\email{pra@imaff.cfmac.csic.es}
\affiliation{Colina de los Chopos, Centro de F\'{\i}sica
``Miguel A. Catal\'{a}n'', Instituto de Matem\'{a}ticas y F\'{\i}sica Fundamental,
Consejo Superior de Investigaciones Cient\'{\i}ficas, Serrano 121,
28006 Madrid, Spain}

\date{\today}

\begin{abstract}
We show that a generalised phantom Chaplygin gas can present a future singularity in a finite future cosmic time. Unlike the big rip singularity,  this singularity happens for a finite scale factor, but like the big rip singularity, it would also take place at a finite future cosmic time. In addition, we define a  dual of the generalised phantom Chaplygin gas which satisfies the null energy condition. Then, in a Randall-Sundrum 1 brane-world scenario, we show that the same kind of singularity  at a finite scale factor arises for a brane filled with a  dual of the generalised phantom Chaplygin gas.
\end{abstract}

\pacs{98.80.-k  95.36.+x}

\keywords{Dark energy, Future singularities, Cosmology with extra dimensions}

\maketitle

\section{Introduction}
In order to describe the current accelerated expansion of our Universe, use has usually been made of a fluid with equation of state $p=w\rho$, where $p$ and $\rho$ are the pressure and energy density, respectively, and $w$ is a parameter which is expected to be fixed by observation. Observational data \cite{Mortlock:2000zu} suggest that $w$ could be less than $-1$ and, in this case, the fluid is called phantom energy. It is well known that when this phantom energy is described as a K-essence field, from which the customary quintessential fluid is nothing but a particular case, the Universe will finish at the so-called big rip singularity \cite{Caldwell:1999ew}, where the scale factor and the energy density go both to infinity and all the structures of the Universe are ripped apart, outside the horizon of each other.
Phantom energy is most generally characterised by its violation of the classical dominant energy condition. The big rip singularity can nevertheless be shown \cite{Gonzalez-Diaz:2003bc,Bouhmadi-Lopez:2004me} to be avoided when the universal fluid is given in terms of a generalised Chaplygin gas (GCG) \cite{Kamenshchik}, even when the dominant energy condition is violated, a case for which the equation of state takes on a more exotic, generally non-linear form. This conclusion can be extended to also encompass the case of the now dubbed dual GCG (DGCG) \cite{nueva}. In this paper it will be seen that, even though the above conclusion can be maintained in the so-named dual version of phantom \cite{Yurov:2006we}, both GCG and DGCG models can predict another kind of future singularity, which might be called a big freeze singularity, as it takes place at a finite value of the scale factor and infinite positive or negative energy density. Therefore, the big freeze singularity is different from the big rip singularity. In fact, the former singularity takes place at a finite scale factor while the latter occurs at an infinitely large radius.

The content of this paper can be outlined as follows. In sec. II we shall consider how in the case of a GCG with positive energy density in a Friedmann-Lema\^itre-Robertson-Walker (FLRW) Universe the big freeze singularity (a singularity type III in the notation of Ref.~\cite{NOT}, see also Refs.~\cite{NOT2,NOT3}) can be obtained. The case for a DGCG occurring in a brane world of the Randall-Sundrum 1 type \cite{RS}, with negative energy density is dealt with in sec. III, where it is shown that also in that case there will appear a future big freeze singularity. We finally conclude and briefly comment on some particular aspects in sec. IV.

\section{Phantom GCG with a future singularity}

The phantom generalised Chaplygin gas  was originally introduced in \cite{Khalatnikov:2003im,Bouhmadi-Lopez:2004me}. It satisfies the same equation of state as GCG \cite{Kamenshchik}, i. e.
\begin{equation}
p=-\frac{A}{\rho^{\alpha}},
\label{chaplygineq}\end{equation}
where A is a positive constant and $\alpha$ is a parameter. If $\alpha=1$, Eq.~(\ref{chaplygineq}) corresponds to the equation of state of a Chaplygin gas. The conservation of the energy momentum tensor then implies
\begin{equation}
\rho=\left(A+\frac{B}{a^{3(1+\alpha)}}\right)^{\frac{1}{1+\alpha}},
\label{rhopcg}\end{equation}
with $B$ a constant parameter. It was noticed in Ref.~\cite{Bouhmadi-Lopez:2004me} that if the parameter $B$ is negative then the perfect fluid with the equation of state (\ref{chaplygineq}) cannot satisfy the null energy condition; i.e. $p+\rho<0$. Furthermore, it turns out that in this case the energy density grows, instead of red-shifting as the Universe expands. This kind of fluid was called a phantom generalised Chaplygin gas (PGCG). Moreover, the authors of \cite{Bouhmadi-Lopez:2004me} (see also \cite{Sen:2005sk}) realised that if the PGCG parameter $\alpha$ is larger  than $-1$ then a FLRW Universe filled with this fluid would escape the big rip singularity because at large scale factors the Universe is asymptotically de Sitter. The authors of Ref.~\cite{Bouhmadi-Lopez:2004me} also pointed out that if  $\alpha$ is smaller  than $-1$ a FLRW Universe filled with this fluid has a maximum finite size but did not work out the consequences in  this case.

\begin{figure}[h]
\begin{center}
\hspace*{-1cm}\includegraphics[width=6cm]{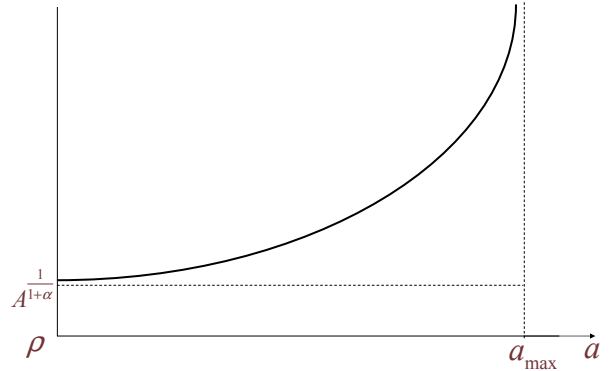}
\end{center}
\caption{Plot of $\rho$ introduced in Eq.~(\ref{rhopcg}) as a function of the scale factor. In this figure, we show the behaviour of $\rho$ for $1+\alpha<0$. A  homogeneous and isotropic Universe filled with this fluid starts its evolution with constant energy density $A^{\frac{1}{1+\alpha}}$ at very small scale factor, therefore there is no big-bang singularity.  The FLRW Universe super-accelerates all the way until it hits a singularity at a constant scale factor $a_{\rm{max}}$ and in a finite future cosmic time. At the singularity the energy density diverges and consequently the Hubble parameter blows up.}
\label{plotrhopcg}
\end{figure}

Here, we will show that a PGCG ($A$ positive and $B$ negative in Eqs.~(\ref{chaplygineq}) and (\ref{rhopcg})) induces a singularity at a finite future cosmic time and for a finite scale factor in a homogeneous and isotropic Universe if $\alpha$ is smaller  than $-1$. This singularity is different from a big rip singularity and we name it a big freeze singularity.
In this case, it turns out that the PGCG energy density is an increasing function of the scale factor (see Eq.~(\ref{rhopcg})). In particular, $\rho$ approaches $A^{\frac{1}{1+\alpha}}$ at very small values of the scale factor ($a\rightarrow 0$) and blows up at a finite scale factor $a_{\rm{max}}$ (see Fig.\ref{plotrhopcg})
\begin{equation}
a_{\rm{max}}=\left|\frac{B}{A}\right|^{\frac{1}{3(1+\alpha)}}.
\end{equation}
Therefore, a FLRW Universe filled with  a PGCG would face a future singularity at a finite radius \cite{footnote1}.
  Moreover, close to the singularity the cosmological evolution can be described as
\begin{widetext}
\begin{equation}
a\simeq a_{\rm{max}}\left\{1-\left[\frac{1+2\alpha}{2(1+\alpha)}\right]^{\frac{2(1+\alpha)}{1+2\alpha}}A^{\frac{1}{1+2\alpha}}
|3(1+\alpha)|^{\frac{1}{1+2\alpha}}(t_{\rm{max}}-t)^{\frac{2(1+\alpha)}{1+2\alpha}}\right\}.
\label{friedmann4d}
\end{equation}
\end{widetext}
For simplicity, we have assumed that the 3-dimensional spatial sections of the FLRW Universe are flat. Hereafter, we shall use units such that $\frac{\kappa_4^2}{3}=1$, $\hbar=c=1$. In Eq.~(\ref{friedmann4d}) $t_{\rm{max}}-t$ measures the cosmic time elapsed since the Universe has a given scale factor  $a$ ($a$ is close to $a_{max}$) till it has its maximum size; i.e.  $a=a_{\rm{max}}$. It was previously pointed out in Refs \cite{Sen:2005sk,Bouhmadi-Lopez:2004me} that a singularity at a finite scale factor can take place in a universe filled by a GCG. Here we have proven that this singularity not only takes place at a finite scale factor but also at a finite future cosmic time.

On the other hand, a FLRW Universe filled with this fluid  is asymptotically de Sitter in the past. In fact,
\begin{equation}
a\simeq a_0\exp\left(A^{\frac{1}{2(1+\alpha)}}t\right),
\end{equation}
where $a_0$ is a small scale factor. Moreover, the FLRW Universe starts its evolution at a past infinite cosmic time where $a\rightarrow 0$ and $p+\rho\rightarrow 0^-$. In fact, the homogeneous and isotropic Universe  super-accelerates; i.e.
\begin{equation}
\dot H=-\frac{3}{2} (p+\rho)>0,
\end{equation}
all the way until it hits the singularity at $a=a_{max}$. In the previous equation, a dot stands for derivative respect to the cosmic time. We recall that the PGCG does not satisfy the null energy condition \cite{Bouhmadi-Lopez:2004me}; i.e. $p+\rho<0$.

In the next section, we will show that the same singularity; i.e. a big freeze, may show up in the brane-world scenario even for a fluid that satisfy the null energy condition. In fact, this can be the case for a dual of the phantom generalised Chaplygin gas.

\section{Dual of a Phantom GCG gas and  RS1 model}

For a given phantom energy model described by a perfect fluid,
we define its dual as a perfect fluid  which satisfies  a similar  equation of state with  $w<-1$ but fulfils the null energy condition \cite{Yurov:2006we}. Here $w$ corresponds to the ratio between the pressure and the energy density of the perfect fluid. This definition of dual has not to be confused with the one used in Ref.~\cite{NOT3}. Despite the very weird property of the dual of a phantom energy model that its energy density is negative \cite{footnote3}, it may supply an alternative to  dark energy models in theories with modified Friedmann equation like the brane-world scenario \cite{review}.  In fact, this might be  the case in Randall-Sundrum model  with a unique brane (RS1) \cite{RS} where the modified Friedmann equation on the brane reads \cite{friedmannbrane}
\begin{equation}
H^2=\rho\left(1+\frac{\rho}{2\lambda}\right).
\label{friedmannrs}
\end{equation}
For simplicity, we consider the flat chart of the brane. In the previous equation $\lambda$ is the brane tension which we assume positive. Consequently, the square of the Hubble parameter in RS1 scenario is well defined for a negative energy density $\rho$ as long as $\rho<-2\lambda$; i.e. as long as the effective energy density of the brane is positive. Moreover, the brane is super-inflating; i.e. the Hubble parameter is an increasing function of the brane cosmic time
\begin{equation}
\dot H=-\frac32 (p+\rho)\left(1+\frac{\rho}{\lambda}\right)>0.
\label{ray}
\end{equation}
It follows that a brane filled with the dual of a phantom energy
is always accelerating; i.e. $\ddot a >0$. The brane is
super-inflating although the dual of the phantom energy satisfies
the null energy density; ie. $p+\rho>0$, because $\rho$ is
negative and smaller than $-2\lambda$. We remind the reader that
the effective energy density of the brane is positive and is is
well behaved (except at the singularity). This is a trademark of
any dual of a phantom energy model.

From now on, we will consider that the brane is filled with a dual of the phantom generalised Chaplygin gas (DGCG) \cite{footnote4}. In this setup, we will show how future singularities may show up at a finite scale factor and in a finite future cosmic time; i.e. a big freeze,  on a  brane filled by this fluid. Indeed, this is the case, for example, for a DGCG, therefore  satisfying the polytropic equation of state (\ref{chaplygineq}), whose energy density reads
\begin{equation}
\rho=\left(A+\frac{B}{a^{-\frac{3}{1+2n}}}\right)^{-(1+2n)},
\label{rhodpcg}
\end{equation}
where $A$ is a negative constant, $B$  a positive constant and $n$ an integer. As can be seen, a commonness of the DGCG is that the parameter $\alpha$ is quantised; in this case $1+\alpha=-1/(1+2n)$.

\begin{figure}[h]
\begin{center}
\includegraphics[angle=270,width=8cm]{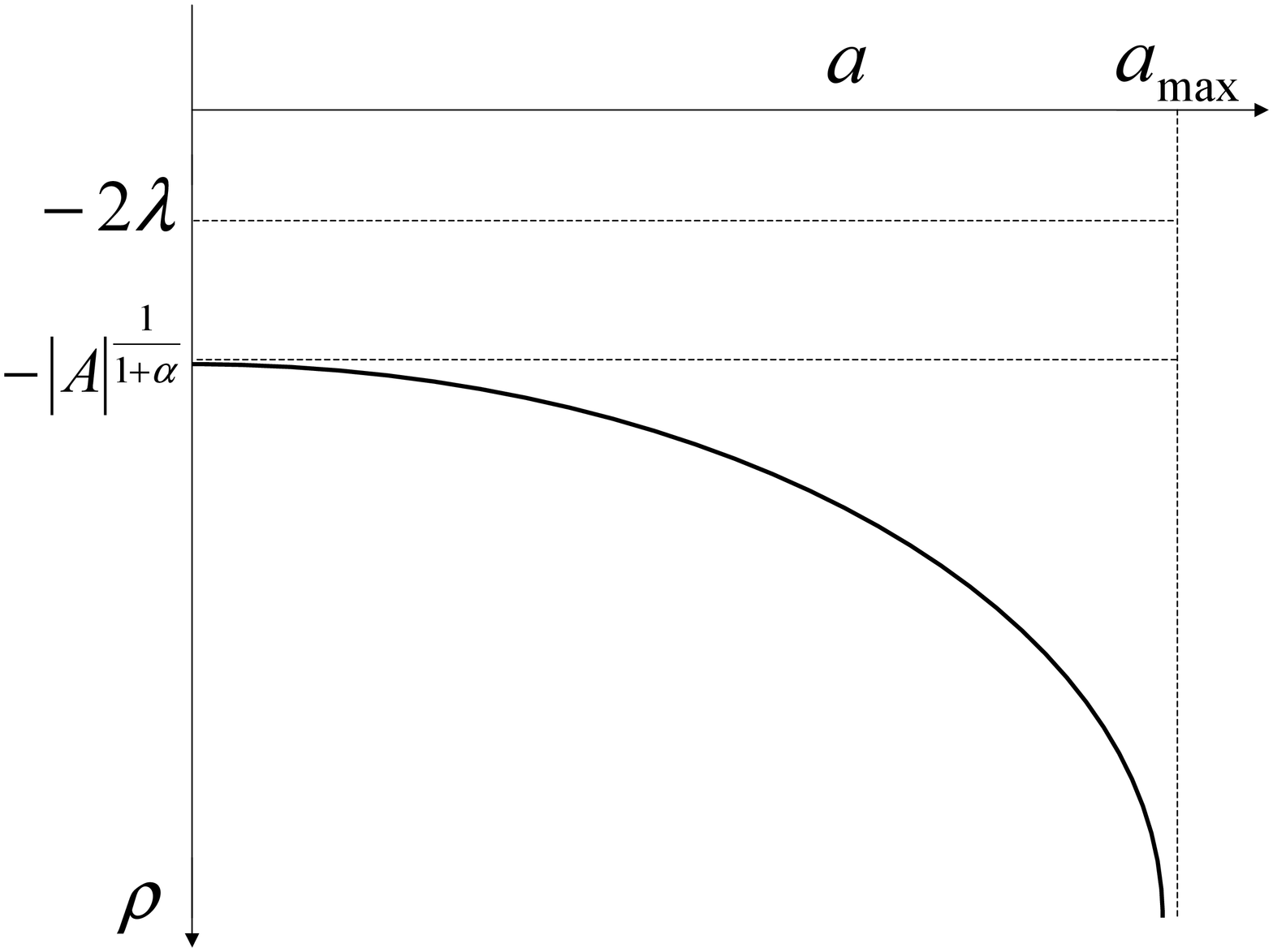}
\end{center}
\caption{Plot of $\rho$ introduced in Eq.~(\ref{rhodpcg}) as a function of the scale factor. In this figure we show the behaviour of $\rho$ ($n$ in Eq.~(\ref{rhodpcg}) is positive) when the maximum energy density is smaller than $-2\lambda$. Consequently, the brane starts its evolution from a vanishing scale factor. The brane super-accelerates all the way until it hits a curvature singularity at $a_{\rm{max}}$ where the energy density blows up in a finite cosmic time $t_{\rm{max}}$.}
\label{rhodpcg1}
\end{figure}

\begin{figure}[h]
\begin{center}
\includegraphics[angle=270,width=8cm]{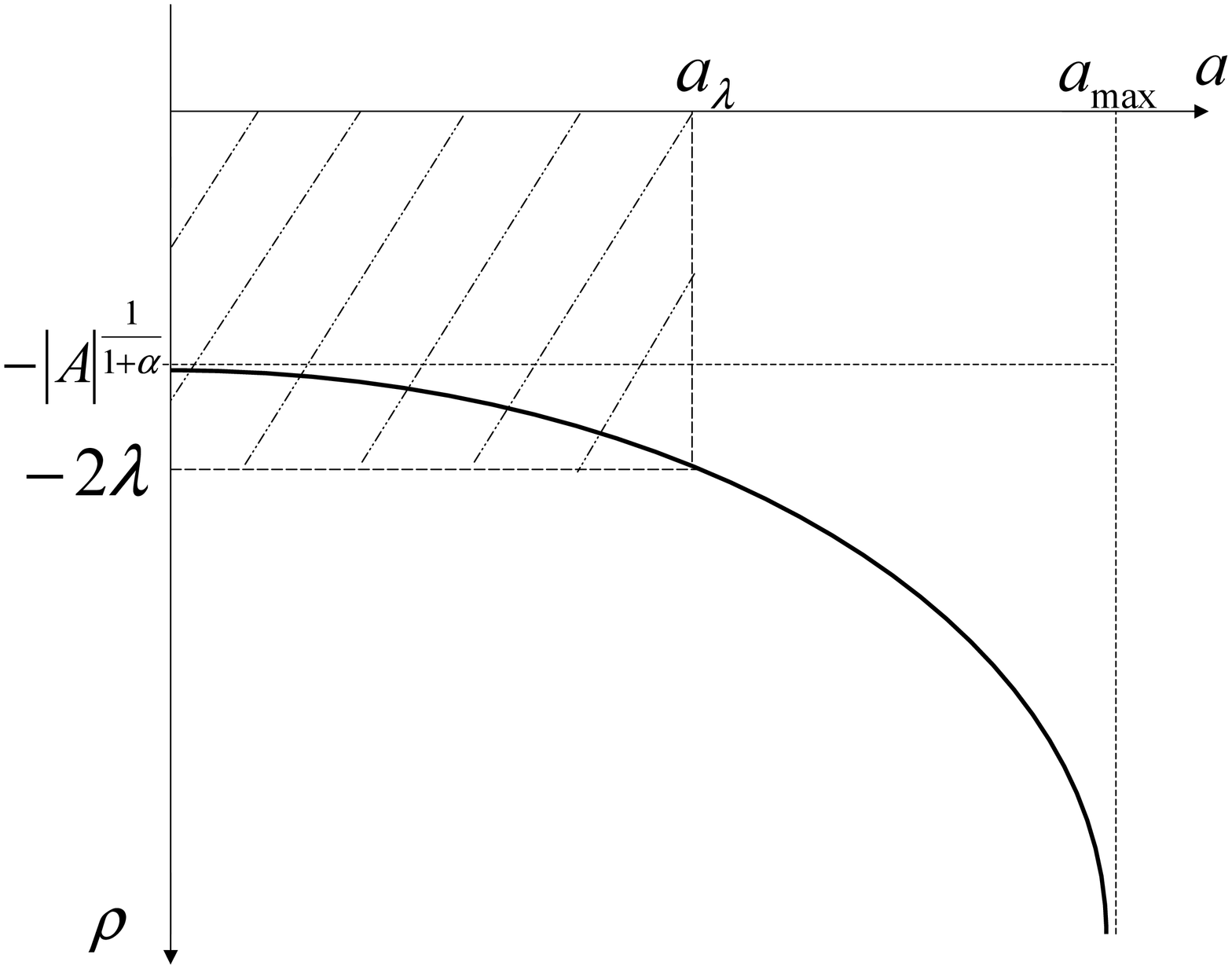}
\end{center}
\caption{Plot of $\rho$ given in Eq.~(\ref{rhodpcg}) as a function of the scale factor. In this figure we show the behaviour of $\rho$ ($n$ in Eq.~(\ref{rhodpcg}) is positive) when the maximum energy density is larger than $-2\lambda$. Consequently, the brane starts its evolution from a non-vanishing scale factor $a_{\lambda}$. The brane super-accelerates all the way until it hits a curvature singularity at $a_{\rm{max}}$ where the energy density blows up in a finite cosmic time $t_{\rm{max}}$. The dashed area corresponds to the region excluded from the evolution of the brane due to the cut off induced by the brane tension on the DGCG energy density.}
\label{rhodpcg2}
\end{figure}

If $n$ is positive, the brane faces a curvature singularity at a finite scale factor $a_{\rm{max}}$ where
\begin{equation}
a_{\rm{max}}=\left|\frac{B}{A}\right|^{-\frac{1+2n}{3}}.
\end{equation}
When the brane reaches its maximum size, the energy density diverges (cf. Eq.~(\ref{rhodpcg})). This feature is schematically shown in Figs.~\ref{rhodpcg1} and \ref{rhodpcg2}. It can be easily checked that close to the singularity the scale factor can be approximated by
\begin{widetext}
\begin{equation}
a(t)\simeq a_{\rm{max}}\left[1-\left(\frac{2}{\lambda}\right)^{\frac{1}{4(1+n)}}|A|^{-\frac{1+2n}{2(1+n)}}(1+n)^{\frac{1}{2(1+n)}}\left(\frac{1+2n}{3}\right)^{\frac{1+2n}{2(1+n)}}(\tilde{t}_{\rm{max}}-t)^{\frac{1}{2(1+n)}}\right].
\end{equation}
\end{widetext}
Here $\tilde{t}_{max}$ is positive and constant. It corresponds to the future finite cosmic time ($\tilde{t}_{max}$) at which the brane hits the big freeze singularity. We would like to stress that the singularity is intrinsic to the brane. In fact, the bulk geometry is well-defined as it corresponds to two symmetric pieces of a 5-dimensional anti-de Sitter space-time split by the brane.

On the other hand, depending on how large is the brane  tension $\lambda$ (for positive $n$), the brane starts its evolution  at a vanishing scale factor or at a finite non-vanishing scale factor $a_{\lambda}$:
\begin{equation}
a_{\lambda}=a_{\rm{max}}\left[1-(2\lambda)^{-\frac{1}{1+2n}}|A|^{-1}\right]^{\frac{1+2n}{3}}.\label{alambda}
\end{equation}
In fact, if the initial energy density of the brane is smaller  than $-2\lambda$; i.e. $2\lambda\leq |A|^{\frac{1}{1+\alpha}}$, the brane starts its evolution with a vanishing scale factor (see Fig.~\ref{rhodpcg1}). In this case, the brane is asymptotically de Sitter in the past and its initial evolution can be approximated by
\begin{eqnarray}
a(t)&\simeq& a_1 \exp(H_0t), \nonumber \\
H_0 &=& \sqrt{|A|^{-(1+2n)}\left(-1+\frac{|A|^{-(1+2n)}}{2\lambda}\right)}
\end{eqnarray}
where $a_1$ is a very small scale factor and $H_0$ is the initial/primordial  Hubble parameter. Therefore, the brane is born at an infinite past cosmic time because  $a$ vanishes at $t\rightarrow -\infty$. Now, if the initial energy density of the DGCG is larger than  the brane tension; i.e. $|A|^{\frac{1}{1+\alpha}}< 2\lambda$, the brane starts its evolution with a non-vanishing radius $a_{\lambda}$ (see Fig.~\ref{rhodpcg2}). Therefore, the brane tension acts as a high energy cut off on $\rho$ \cite{footnote2}. This feature implies that $\lambda$ forbids an infinite past of the brane. Indeed, close to the initial scale factor $a_{\lambda}$, the cosmological evolution can be described as
\begin{widetext}
\begin{equation}
a\simeq a_{\lambda}\left\{1+\frac34\frac{a_{\lambda}}{a_{\rm{max}}}(2\lambda)^{\frac{2(1+n)}{1+2n}}
|A|\left[
1-(2\lambda)^{-\frac{1}{1+2n}}|A|^{-1}\right]^{\frac{2-2n}{3}}(t-t_{\lambda})^2\right\}.
\end{equation}
\end{widetext}
where at $t=t_{\lambda}$, the brane radius coincides with $a_{\lambda}$. In summary, we have shown that a DGCG for $n$ positive (see  Eq.~\ref{rhodpcg}) induces a future singularity at a finite future cosmic time $\tilde{t}_{\rm{max}}$ and for a  finite radius of the brane $a_{\rm{max}}$. On the other hand,  we have also shown that depending on how large is the brane tension, the brane starts its evolution with a vanishing radius at an infinite past or with a finite radius $a_{\lambda}$ at a finite past $t_{\lambda}$ (as measured by the cosmic time of the brane). At this respect, we would like to emphasise that we are considering only the phase where the brane expands. It is worth noticing that the expansion phase is preceded by a collapsing phase with negative Hubble parameter. It can be also seen that the latter phase starts with a past freeze singularity.

\begin{figure}[h]
\begin{center}
\includegraphics[angle=270,width=8cm]{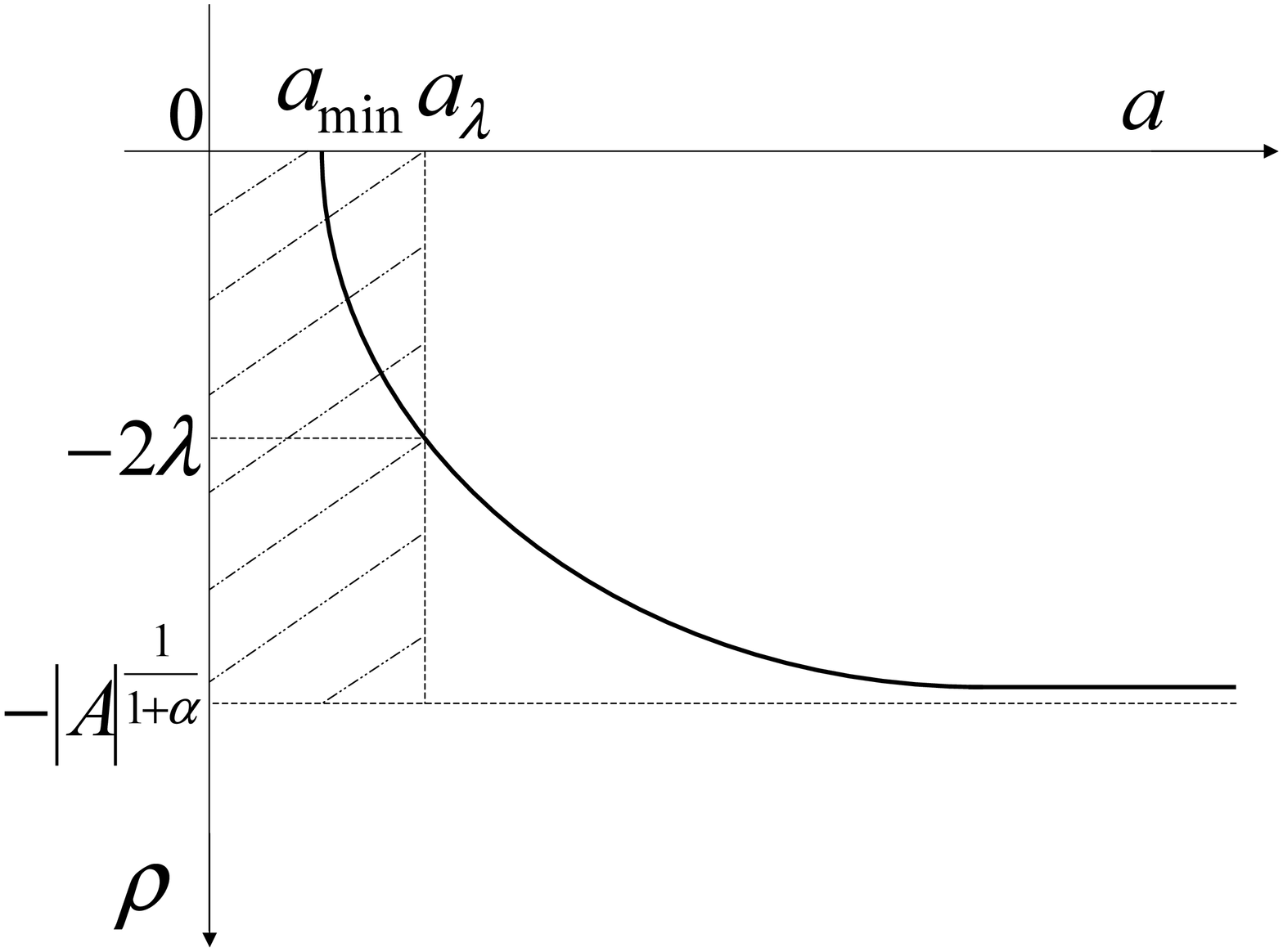}
\end{center}
\caption{Plot of $\rho$ introduced in Eq.~(\ref{rhodpcg}) as a function of the scale factor. In this figure, we show the behaviour of $\rho$ ($n$ in Eq.~(\ref{rhodpcg}) is negative) when the minimum energy density is smaller than $-2\lambda$. Consequently, the brane starts its evolution from a non-vanishing scale factor $a_{\lambda}$. The brane super-accelerates all the way until it becomes asymptotically de Sitter in an infinite future cosmic time (of the brane).}
\label{rhodpcg3}
\end{figure}

Before concluding this section, we describe briefly the evolution of a brane filled by a DGCG whose energy density is given by Eq.~(\ref{rhodpcg}) for $n$ negative (for more details see Ref. \cite{nueva}). First of all, the minimum energy density of the  DGCG must be smaller than $-2\lambda$ (see Fig.~\ref{rhodpcg3}); i.e. $2\lambda< |A|^{-(1+2n)}$ ,  otherwise the brane cannot have a Lorentzian evolution. Under this assumption, it turns out that the brane is asymptotically de Sitter in the future despite that the DGCG energy density is negative (see Fig.~\ref{rhodpcg3}). At this respect, we recall that the modified Friedmann equation on the brane  is quadratic on the total energy density of the brane (cf. Eq.~(\ref{friedmannrs})). In addition, the effective energy density of the brane is positive. The brane tension acts again as a high energy cut off on the DGCG energy density. Consequently, the brane starts its evolution at a non-vanishing scale factor $a_{\lambda}$, previously introduced in Eq.~(\ref{alambda}). Therefore, the brane tension excludes those radii of the brane such that $a_{\rm{min}}<a<a_{\lambda}$, where the energy density vanishes at $a_{\rm{min}}$. This cut off effect implies that the brane starts its evolution at a finite past cosmic time as measured by an observer confined on the brane.

\section{Discussion and conclusions}
We have seen how GCG and DGCG could generate a new big freeze singularity (a singularity type III in the notation of Ref.~\cite{NOT}) which would be associated with a different way for the Universe to finish along a doomsday. At this doomsday the Universe would be infinitely full with phantom energy while it would have a finite size, in contrast with a doomsday {\it\`{a} la} big rip. Thus, nothing in the Universe could move around, producing a freezing for all the eternity.

Close to a big rip singularity, the classical space-time breaks down and therefore a quantum analysis at large scale factors is required \cite{Dabrowski:2006dd}. We expect similarly the quantum effects to  be also important close to a big freeze singularity and consequently at intermediate values of the scale factor. In particular, the kind of big freeze singularity we have found  in this paper needs  an analysis of the GCG from a quantum point of view \cite{Bouhmadi-Lopez:2004mp}. We leave  this interesting issue for a future work.

\section*{Acknowledgements}

MBL acknowledges the support of CENTRA-IST BPD (Portugal) as well as the FCT fellowship SFRH/BPD/26542/2006 (Portugal). The initial work of MBL was funded by MECD (Spain). MBL is also thankful to IMAFF (CSIC, Spain) for hospitality during the realisation of part of this work. PMM gratefully acknowledges the financial support provided by the I3P framework of CSIC and the European Social Fund. This research was supported in part by MEC under Research Project No. FIS2005-01181.

\vspace*{1.5cm}

\begin{thebibliography}{99}
\bibitem{footnote1} The same behaviour can be found when one considers $A$ negative,
$B$ positive and $(1+\alpha)^{-1}=2n<0$ ($n\in\mathbb{N}$)
\cite{nueva}.
\bibitem{footnote4}
A clarification is in order before proceeding. If we apply our model to the current epoch of the universe, the constraint on the upper limit of the energy density of the DGCG, $\rho$, is modified by the presence of dark matter. More precisely, $\rho< -\rho_m-2\lambda$, where $\rho_m$ corresponds to the energy density of dark matter. This condition insures the positiveness of the square of the Hubble parameter and the super-acceleration of the brane (see Eq.~(\ref{ray})). We have ignored the dark matter energy density because we are interested in analysing the future big freeze singularity and when this event takes place the red-shifting dark matter energy density is negligible. On the other hand, we expect our model to be valid  from low red-shift onwards (towards negative red-shift). Precisely, for this reason we do not expect our model to be valid at very early time when nucleosynthesis took place and where the dual phantom energy should be expected to be sub-dominant or actually absent.
\bibitem{footnote2} Whenever we refer to a high energy effect, we mean an effect at
small scale factors.
\end{thebibliography}
\end{document}